# Alternative Vision of Living with IoT

**EunJeong Cheon**
Indiana University
Bloomington, IN 47408, USA
echeon@indiana.edu

## Abstract
In this submission, I discuss my research on values, norms and practices of subcultures formed as an "alternative" to the dominant way of life. In particular, I explore how the Internet of Things (IoT) or intelligent agents relates to alternative forms of interaction or be understood and reconstructed through alternative concepts or frameworks. For the past three years I have been conducting fieldwork on communities pursuing alternative lifestyles. This work considers how those alternative lifestyles may contribute to an understanding of objects, spaces in future smart home. Through my fieldwork and research through design, I hope to offer an alternative vision to living with IoT and envision future domesticity in a unique and even groundbreaking way.

## ACM Classification Keywords
H.5.m. Information interfaces and presentation (e.g., HCI): Miscellaneous;

## Introduction
Over the recent years there has been increasing interest in alternative ways of living that bring about some changes in the principles and patterns of everyday life, for example, by simply adopting a new dietary habit (e.g., a low carb high fat diet) or engaging in zero waste practices at home. As we are exposed to or embrace various lifestyles and practices, it becomes increasingly important to understand how technologies may be intermingled with societies.

My research lies at the intersection of subculture and everyday life practices surrounding the technologies we develop, focusing on how new technologies can be implemented in emerging subcultural practices which must consider their relationship with mainstream society. To examine this, I look at alternative values, norms and practices of subculture groups, and explore how the Internet of Things (IoT) or intelligent agents can manifest alternative forms of interaction or be understood and reconstructed through alternatives concepts or frameworks. Through my fieldwork and

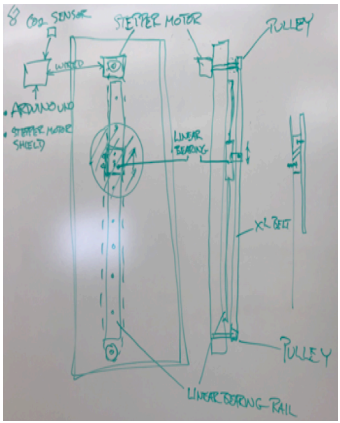

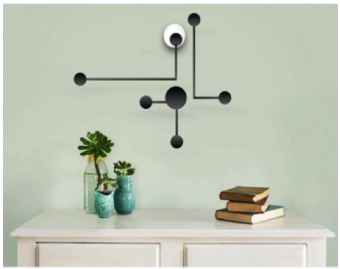

Figure 1: Sketching and prototyping process of Lines & Dots (above), and expected look and feel of Lines & Dots (below).

.

design practices, I hope to develop alternative visions of living with IoT and intelligent agents.

## Fieldwork on Alternative lifestyles

For the past three years I have been conducting fieldwork on self-identified minimalists: from zero-waste practitioners to tiny house enthusiasts. My case studies have taken place in three major US cities and two cities in Korea. Minimalists have grown in popularity in recent years. They pursue the value of minimalism by getting rid of much of modern life's clutter. Despite the many motivations and variations in incorporating minimalism into daily life, all seem to share a central practice: re-evaluating what we possess and eventually having less.

This project explores how alternative values and practices may inform designing new technologies for home and domesticity. In [1], we describe minimalists' unique perspectives with regard to space, which they have gradually honed through their de-cluttering practices. For a minimalist, *empty space* is to be lived-in, visible, and interactive. This understanding forces us to reconsider previous understandings of the "interactivity" of a design; empty space is no longer a blank canvas waiting to be filled, it is a conscientious, aesthetic choice.

One of my crucial discoveries has been the importance of the home for a minimalist as a reflection of their values. When committed to a minimalist lifestyle, it is inevitable that minimalists will need to negotiate with others regarding objects in and outside the home. Such negotiation makes the boundaries of their home *porous* when they are forced to share space with unwanted objects. I have been developing *porous boundary* as an analytic lens to understand how artifacts, coupled with their values, are preserved or resisted through the boundaries of a place [2].

Current issues minimalists struggle with include the negotiation between empty space and the cluttering in the home caused by "smart" technology. For example, placing several Amazon Echos around the home for seamless connections to other home technologies can be viewed as both an act of minimalism and an act of materialism. This calls for rethinking our market-oriented design logic where adding more new things makes your home smarter. Regarding this, minimalist downsizing practices resonate with alternative design approaches such as undesigning [5], Fry's elimination design (de and re-materialization) [3], and Tonkinwise' disowning design strategy [6].

As the next step (partially illustrated in Fig2), I design IoT prototypes using a research through critical design [4] process—this combines research through design with critical inquiry. The prototypes would enable in-depth exploration with the study participants on their "assumptions, values, ideologies, and behavioral norms" (p.303).

## Research Through Prototyping IoT

I am currently developing different design prototypes in parallel and plan to deploy them in minimalists homes, tiny houses and/ or shops run with zero-waste practices. The prototypes that I have explored are 1) Lines & Dots, which visualizes the big picture of the IoTs at home, 2) Food Trace: a supply chain tracker of bulk food at stores, which encourages package-free shopping by offering the supply chain data of products, and 3) Deodorizing Machine, a smart fragrance diffuser that is activated dependent on the time durations required by cooking appliances.

As one example, Lines & Dots (Fig 1) is a wall-mountable IoT. It is intended to provide a bird's eye view of IoT technologies operating across the different spaces of a house. The shapes of the lines and dots abstractly represent the home's floor plan. Small dots signify rooms or separate spaces, and are connected with lines that stretch toward the center of the home. The big dot in the center marks the home's central space, in the most cases this is the living room. All dots are embedded with NFC chips that are unique to each room and interact with the technologies enclosed in that room. They detect the number of IoT devices and lighten up a NeoPixel attached behind the dot. With its distinct level of brightness and movements, this prototype offers the big picture of the technologies breathing in our living spaces and amplifies their presence.

## In the workshop

In this workshop I would like to discuss the following three questions with my fellow participants:

1) How could everyday artifacts become a part of the IoT ecology? (E.g., how could everyday objects be augmented into IoT devices? How can we design the future IoT without pre-defining smart objects? How might non-technological objects be involved in the current ecology of technological objects?)

2) How could a "smart" or an "intelligent" agent be embodied by multiple forms? (E.g., how could my "Alexa" be transferred from one device to another, or even several others? How could such an intelligent agent "co-exist" in more than one device?)

3) How could IoT be space-sensitive? (E.g., how could IoT intentionally be disconnected in certain spaces, such as the bedroom? How could future IoT work in mobile conditions or in spaces with porous boundaries?)

4) (This is speculative.) How might the future direction of IoT help us move toward an alternative economy (e.g., from a linear economy to a circular economy)?

Furthermore, I would like to discuss what alternative methods or frameworks – even provocative ones – our community could leverage to tackle to these questions.

on Human Factors in Computing Systems (CHI '12)*: 957–966. http://doi.org/10.1145/2208516.2208540

6. Cameron Tonkinwise. 2013. Design Away. *Design philosophy politics*: 1–14. Retrieved May 27, 2018 from https://www.academia.edu/3794815/Design_Away_Unmaking_Things